\documentclass[12pt]{article}
\title{\bf Hamiltonian  two-body system in special relativity}

\author { Philippe Droz-Vincent\\[2mm]LUTH\\
Meudon \footnote{Observatoire de Paris, CNRS, Universit\'e Paris Diderot,
5  place Jules Janssen,   92195  Meudon, France   }}

\date{ 
      }

\newcommand  {\eeq}{\end{equation}}
\newcommand  {\beq}{\begin{equation} }
\newcommand  \half {  {1 \over 2} } 
\newcommand  \const{ \rm const. }
\newcommand  {\ytil}{\widetilde y}
    
\newcommand{\delhat}{\widehat \delta} 
\newcommand  {\ptil}{\widetilde p}     
\newcommand {\hatx}{\widehat x} 

\newcommand {\ztil}{\widetilde z}
\newcommand {\rtil}{\widetilde r}

\newcommand  {\noi}{\noindent}  
\newcommand  {\jok}{A}
\newcommand  {\joka}{a^2}    
\newcommand  {\jokb}{b^2}
\newcommand  {\disp}{\displaystyle}
\newcommand  {\mun}{ {\mu  \nu } }

\newcommand {\sig}{   \sigma   }
\newcommand {\vareps}{   \varepsilon  }
\newcommand {\Sig}{   \Sigma   }
\newcommand {\alp}{\alpha} 

\newcommand {\phat}{\widehat p}

\newcommand{\del}{\delta}

\newcommand{\gam}{\gamma}
\newcommand{\lam}{\lambda} 
\newcommand{\Lam}{\Lambda} 
\newcommand{\ome}{\omega}
\newcommand{\Ome}{\Omega}

\newcommand{\calm}{{\cal  M}}

\newcommand{\dron}{\partial}   

\newtheorem{prop}{Proposition}
 \newtheorem{theo}{Theorem}

\newcommand{\beprop}{\begin{prop}}
\newcommand{\betheo}{\begin{theo}}
\newcommand{\enprop}{\end{prop}}
\newcommand{\entheo}{\end{theo}}
\begin{document}
\maketitle \abstract{We consider an isolated  system made of two pointlike bodies interacting at a distance in the nonradiative approximation.    Our framework is the covariant and  {\em a priori} Hamiltonian formalism of "predictive relativistic mechanics", founded on the equal-time condition. The center of mass is rather a center of energy. Individual energies are separately conserved and the meaning of their positivity is discussed in terms of world-lines.  Several  results derived decades ago under restrictive assumptions are extended to the general case.  Relative motion has a structure similar to that of a nonrelativistic one-body motion in a stationnary external potential, but its evolution parameter is generally not a linear function of the center-of-mass time, unless the relative motion is circular (in this latter  case the motion is periodic in the center-of-mass time).   
Finally the case of an extreme mass ratio is investigated. When this ratio tends to zero the heavy body  coincides  with the center of mass provided that a certain first integral, related to the binding energy, is not too large. 

 }

\bigskip

\section{Introduction}
Classical relativistic dynamics of pointlike bodies has a long story; without claiming to be exhaustive let us mention the Wheeler-Feynman (WF) electrodynamics \cite{WF} based upon the Fokker action~\cite{Fokk}, the three forms of dynamics (front form,  point form, instant form) advocated by Dirac~\cite{dirac},   and  after the discovery of a famous No-Interaction theorem~\cite{Currie}, various efforts made in order to circumvent it; for instance Predictive Mechanics~\cite{droz}\cite{predico}, the Singular Lagrangian method~\cite{longhi} and  Constraint Dynamics~\cite{tod}.  In the last decade there were a few papers  along the lines of WF~\cite{uryu}  and  also the  work  carried out by  Lusanna {\em et al.} \cite{lusanetal} in order to give a covariant status to the instant form.  

\noi Beside several $n$-body generalizations, most progresses have been devoted on the two-body problem, which is our present subject.

A first point was the possibility of actually having (unlike WF electrodynamics) second-order differential equations describing the  motion of a {\em finite number} of degrees of freedom interacting at a distance. According to this view,  the field that carries interaction is supposed to be eliminated, the space of initial data has a finite number of dimensions and a point in this space uniquely determines the subsequent motion~\cite{droz} \cite{predico}.

\noi
A second point was about a Hamiltonian formalism. We have in mind conservative mechanics: radiative corrections are neglected and  we focus on isolated systems, characterized by a finite number of degrees of freedom and by Poincar\'e   invariance.  
Under these conditions the no-interaction theorem~\cite{Currie} excludes the possibility of demanding that the physical positions be canonical throughout phase space (as was always done in classical mechanics). Relaxing this requirement implies an  arbitrariness which  can be  removed by imposing a relationship  between  physical positions and canonical variables  on some  submanifold $\Sig$. 

\noi  In the {\sl a priori Hamiltonian} approach of predictive mechanics~\cite{repor}, 
  manifest covariance is realized with help of considering degrees of freedom that are geometrically natural but redundant (if compared to the classical situation).
Positions and momenta are four-component objects, phase space has sixteen independent dimensions and we employ a multitime formalism 
(in a  different spirit Todorov proposed to focus on  the physical degrees of freedom only, in another  formulation of dynamics which is covariant as well but includes 
 first-class {\sl constraints}~\cite{tod}). 

After discussing  other possibilities, we  put forward a  natural prescription:    physical and canonical  positions must coincide when both physical positions are simultaneous with respect to the rest frame of the system~\cite{repor}. In this approach a submanifold  $\Sig$ is obtained by selecting the configurations where the physical  coordinate times $x_1^0 , x_2 ^0$ are equal in the center-of-mass frame.  An advantage of this {\sl equal-time prescription} is to permit the contact with the constraint formalism, as shown in detail by L. Lusanna~\cite{luca}.
 
\medskip
The two-body problem immediately suggests these  important issues: center of mass and relative motion. In this  
 article we intend to study their properties in the context  of relativistic dynamics, aiming at a method  in order to simplify  the  determination of  the world lines.  

A few exploratory results that we derived in the past~\cite{annihp} \cite{cras} have remained fragmentary, as most of them  have been obtained with help of restrictive assumptions  concerning the shape of the interaction potential. The present work is free of such limitation, aiming at the possibility of dealing with realistic interactions; such interactions  cannot offer the simplicity of the academic models we considered long time ago: see for instance the unavoidable $P^2$- dependence of an electromagnetic two-body potential proposed by Jallouli and Sazdjian \cite{sazjallou}, in the framework of  relativistic {\em quantum} mechanics.

Before we focus on its  equal-time  version, we shall sketch the main lines of the
 {\em a priori} Hamiltonian approach in general.

\noi
In the next section we present the isolated two-body systems and introduce the center of mass.

\noi  In the rest of the paper we  consider unipotential models (for these systems the  individual energies are separately conserved).

\noi
In Section 3  we  study the evolution of the canonical variables, irrespective of their relationship to physical positions.  
Later, in Section 4,  we  focus on the equal-time prescription and discuss  a strategy for the determination of world-lines; the importance of circular orbits is emphasized.
In 5  we consider  the case of an extreme mass ratio; a toy model is presented in Section 6, and  Section 7 is devoted to  concluding remarks.

\bigskip
\noi The velocity of light is taken as unity, except in the Appendix.
When no confusion is possible, tensor indices are omitted and the contraction dot is employed also for tensors, for instance $\   J \cdot P  \   $ stands for the vector
$ J^{\alp \mu}\   P_\mu $.

\subsection {The  {\em a priori} Hamiltonian formalism}
    
The canonical coordinates are   $q_a ^\alp , p_{b  \beta} $ in the sixteen-dimensional phase space; $q_1 , q_2 $ are
 points in Minkowski space $\   \calm  \quad  $  and  $\     p_1 , p_2  \    $ are four-vectors.      

\noi The symplectic form 
  $ dq_1 ^\alp \wedge dp_{1 \alp}  +   dq_2 ^\alp \wedge dp_{2 \alp}    $               corresponds to the Poisson brackets
\beq   \{ q_a ^\alp  , p_{b  \beta}  \}   =   \del _{ab} \    \del ^\alp _\beta       \label{poiss} \eeq 
Note that 
the  particle labels $a,b,c $ are {\em not} summed over when repeated.

\noi
The Hamiltonian  equations of motion 
\beq  {\dron q_a  \over \dron \tau _b }  =  \{ q_a  , H_b \}    \qquad  \quad
      {\dron p_a  \over \dron \tau _b }  =  \{ p_a  , H_b \}
\label{hamil} \eeq
involve two Hamiltonian generators $H_1 , H_2$ submitted to the {\sl predictivity condition}
   $$  \{ H_1 , H_2 \} = 0               $$
The  two evolution parameters $\tau_1 , \tau _2 $   are suitable generalizations of the proper times (normalized to the masses).
 Our notation is choosen such that the generators can be identified as the half-squared masses; this allows to consider the masses as first integrals,   redeeming  the fact that phase space has redundant degrees of freedom.

\noi For instance, in the trivial  case of two free particles one is left with 
$\disp  2H_a = p_a ^2$, the physical positions reduce to $x_a = q_a$ and the evolution parameters are just $\tau _a = s_a / m_a$, where $s_1 , s_2$ are the arc lenghts.

\noi  For {\em interacting} particles the  Hamiltonians involve additional terms $\   V_1 , V_2 \    $referred to as "potentials".
In general $q_1 , q_2$   differ from the physical positions, in fact $x_1 ,  x_2$ are determined by the partial differential equations
\beq \{  x_a , H_b \} = 0    \qquad \qquad  {\rm for}     \qquad a \not=  b
\label{positeq} \eeq
and some reasonable initial data. 
Solving these equations 
 provides  a correspondance between physical and canonical coordinates, say 
     $$x,v   \longleftrightarrow       q,p  $$
which, inserted into the  solutions of  system (\ref{hamil}), ensures that 
 $\  x_1 \  $ is a function of $\tau _1$ only and  
$\  x_2 \  $ is a function of $\tau _2$ only, in other words (\ref{positeq})
ensures the existance of one-dimensional world-lines~\cite{repor}. This point is easily checked by introducing on phase space the Hamiltonian vector fields $X_1 , X_2$
 defined by  
\beq  X_a\    f = {\dron  f \over  \dron   \tau _a }     =   \{ f , H_a \}   \eeq
 for all phase-space function $f$. 
The  predictivity condition  corresponds to the vanishing of      the Lie bracket $ [X_1 , X_2 ]$.   
Although the analytic shape of the Hamiltonians yields some information about first integrals, symmetries, etc, the physical meaning is fixed only once a solution to (\ref{positeq}) has been specified.  At this stage phase space is identified with  the bundle product $ T( \calm ) \times  T (\calm )$.


\medskip
\noi Notation.  $\disp  \     (H_a )_{\rm free } =  \half p_a ^2 \    $, they generate  
$( X _a ) _{\rm free }$.

\noi   When fixing numerical values $\half m_1 ^2 $ and  $  \half m_2 ^2 $ to the Hamiltonian  generators,  it is convenient to set
 \beq  \mu = \half (m_1 ^2 + m_2 ^2 ) , \qquad \quad
 \nu = \half (m_1 ^2 - m_2 ^2 )     \label{munumass}           \eeq 
which amounts to 
\beq   m_1 m_2  =   \sqrt{ \mu ^2 - \nu ^2} , \qquad \qquad
   ( m_1 + m_2 )^2    =  2 \mu  +  2  \sqrt{ \mu ^2 - \nu ^2}  
   \label{remember} \eeq

\section{Isolated two-body systems}

The Lie algebra of the Poincar\'e group is spanned  by 
$ P_ \alp ,  J_ \mun  $.
We separate  external from internal   variables by setting 
\beq  P= p_1 + p_2    \qquad \quad  Q= \half (q_1 + q_2 )              \eeq
\beq  y = \half (p_1 - p_2 )       \qquad  \quad   z = q_1 - q_2      \eeq
Remark $\     $  in general  $\   Q \   $ {\em is not} the center of mass~\footnote{even for free particles it is center of mass {\em only when} the masses are equal.}.
 
\noi  Other definitions of $Q$  and $y$ (as conjugate to P and z respectively) were possible, but the advantage of ours is that it does not require to  {\em a priori}  fixthe numerical values of the constituent  masses. 

\noi The standard Poisson brackets that do not vanish can be re-arranged as follows
\beq  \{ Q^\alp , P_\beta  \} = \{  z^\alp , y _\beta  \}  =  \del ^\alp _\beta     \label{canon} \eeq
and we can write
$$  J =   Q \wedge P    + z \wedge y   =  q_1 \wedge  p_1  +  q_2  \wedge  p_2  $$
Naturally $H_1 , H_2$ are supposed to be Poincar\'e invariant (vanishing Poisson brackets with $P$ and $J$) and the initial conditions for solving  (\ref{positeq}) must be invariant as well.

\medskip
 From   (\ref{canon}) we can derive several useful formulas; let us list some of them. 
Of course  $ \{ Q^\mu ,  P^2 \}    =  2  P^\mu $ and
$$    \{ Q^\mu  ,  P_\alp  P_\beta  \} =    \del ^ \mu  _\alp  P_\beta
   +  P_\alp  \del ^\mu  _\beta  $$ 
so if we define         the projector orthogonal to $P$
\beq
  \Pi = \eta -  {P  \otimes  P   \over  P^2 }   \label{defPi}     
\eeq
     we get
\beq    \{ Q ^\mu  ,  \ztil ^2  \} =  - 2 \ztil ^\mu { P \cdot z  \over P^2 }
\label{Qmuztil}          \eeq
with this notation that the {\em tilde means an application of} $\Pi , \   $ say
$$ {\widetilde \xi}^\alp =  \Pi^ {\alp  \beta}    \   \xi _\beta,
 \qquad \quad  \forall  \xi     
$$
the r.h.s. of (\ref{Qmuztil})  is  orthogonal to  $P$; it follows that 
\beq    \{ Q \cdot P  ,  \ztil ^2  \} =      0
\label{QPztil}   \eeq


\medskip
\noi On the other hand we compute

\beq   \{ Q \cdot  P ,  \     {P  \otimes  P   \over  P^2 }   \} =  0   
\label{QPproj}        \eeq
in other words    $Q \cdot  P$ has a vanishing    Poisson bracket with the projector
$\disp  \Pi $.   
 Moreover we easily check that
\beq      \{ Q \cdot  P  ,   { (y \cdot P )^2   \over P^2 }  \}       = 0
\label{yP2surP2}    \eeq
                                                                   
\noi
It is obvious that $ \{  Q \cdot P  ,   y^2 \}  $ vanishes and we can apply the above formula to the identity
$$ \ytil ^2 =  y^2  -      { (y \cdot P )^2   \over P^2 }  $$
  which yields
\beq  \{ Q \cdot P  ,  \ytil ^2  \} = 0   
      \label{QPytil}                      \eeq
For the relative variables note that  the spatial piece  of the one has a vanishing bracket with the time piece of the other (its conjugate):
\beq    \{  \ztil ,\    y \cdot P  \} =   \{  \ytil ,\    z \cdot P  \}  =0
                        \label{esptmp}       \eeq
Similarly it stems from   (\ref{QPproj}) that 
\beq  \{ Q \cdot P ,  \ztil ^\alp \}  =     \{ Q \cdot P ,  \ytil ^\alp \}  = 0 
      \label{QPzytil}       \eeq

\bigskip

Of course  we  already know ten first integrals, namely  $P_\alp,  J_\mun $. We obviously have that
\beq  {  \widetilde{ P \cdot J } \over  P^2 } =
{ z  \cdot P  \over   P^2   }  \     \ytil     -  \widetilde{Q}  -  
   { y \cdot P   \over  P^2 }  \     \ztil           \label{Qtil}                \eeq      

\bigskip
\subsection{ Center of Mass}

\noi The possibility to define a center of mass, using linear and angular momenta
as basic ingredients, has been  known long time ago~\cite{pryce}.

\noi
In a previous work \cite{JMP1996} we proposed this canonical definition for the 
components of the center of mass

\beq  \Xi  =   {J \cdot P  \over  P^2} +  ({ P \cdot Q  \over P ^2 }) \   P
   \label{defcdm}         \eeq
or equivalently
\beq     \Xi  = Q +  ({y \cdot P \over  P ^2})   z   -
                       ({z \cdot P \over P^2  }) y
\label{vardefcdm}       \eeq
This  can be  transformed again, if we notice that 

\noi $\disp {P^2 /2} \pm y \cdot P =  P \cdot p_1 , \   ( {\rm resp. \    }  P \cdot p_2 )$, we get
\beq 
\Xi = {P \cdot p_1 \over P^2 } q_1  +    {P \cdot p_2 \over P^2 } q_2
  -  {P \cdot z     \over P^2} y
\label {varvarcdm}  \eeq
Formula (\ref{vardefcdm})  entails 
\beq   \Xi  \cdot P  =  Q \cdot  P     \label{tempXiQ}  \eeq



\noi 
In  (\ref{defcdm}) the only quantity  which is not a constant of the motion is 
   $ \    ( {P \cdot Q  /  P^2 } ) \         $,
it has the same dimension as     $\tau _1 , \tau _2 $.
 In contrast setting~\footnote{ definition of $ T$ has a different dimension in \cite{JMP1996} , but the same as here in  \cite{cras} ) }.
  $$   T =  {P \cdot Q \over  |P| }  $$ 
we give to  $T$ the dimension of lenght (and {\em time}  since $c=1$).
 One easily computes 
$$ \{    {(J \cdot P )  ^\alp  \over  P^2}  , \   P_\beta  \}  = \Pi ^\alp _\beta 
$$ 
whence we derive the relations
\beq   \{ \Xi ^\alp , \   P_\beta  \}  =  \del ^\alp _ \beta        \qquad 
\quad     \{ \Xi ^\alp , \  { P^2  \over 2}  \}   =   P^\alp 
   \label{canoncdm}        \eeq

\noi  Owing to the constancy of $J_{\mun}$ and $P_\alp$, we can fix these quantities, say in particular
$$ P_\alp = k_\alp  ,             \qquad  \quad  {\rm timelike  \   vector  }              $$ 
$$  {\rm  define}  \qquad  \qquad  \               k^\alp  k_ \alp  =  M^2       $$
Then  we  see that formula       (\ref{defcdm})         defines the coordinates of a point which {\em moves on  a straight line when both $\tau _1 , \tau _2 $ run freely}  and independently  from 
$- \infty$ to $+ \infty$, and the direction of this line is given by $k^\alp$.

\noi  Equation (\ref{defcdm}) becomes  
$ \disp \      \Xi ^\alp = 
 {\rm Const.} +  ({P \cdot  Q \over M})\     {k^\alp \over M}$.

\noi   We can write  $ \disp  {d \Xi ^\alp \over  dT } = {k ^\alp  \over  M } $ and consider  respectively  $T $ and  $ T/M$ as the {\em proper time }  and the  {\em evolution parameter } of the center of mass.
 
 Similarly  equations (\ref{canoncdm} ) supplemented by the trivial observation that $\disp  \{ P ,  \half P^2  \}$  vanishes, can be viewed as canonical equations of motion for $\Xi $,  generated by the one-body Hamiltonian    
$\   \half P^2 \    $.


 Note however that in  (\ref{defcdm}) the components of $\Xi$ {\em do not} commute among themselves; a similar situation was already encontered by Pryce~\cite{pryce}.

\bigskip
 From now on we shall focus on {\sl Unipotential Models} characterized  by the same interaction term for  both particles, say $V_1 = V_2   =V$.
Such models permit to  write down explicit forms of the Hamiltonians.  
Moreover it seems that they  are still general enough for dealing with most realistic interactions.

\section{Evolution of the canonical variables }
For the moment we postpone the physical interpretation; we are interested in the evolution of the canonical variables 
 in terms of the  parameters   $\tau _1 ,  \tau _2 $, more generally we are concerned with  statements that hold true regardless to the prescription used for solving the position equations  (\ref{positeq}). In this section, the only thing we assume about this prescription is Poincar\' e invariance.

At this stage the separation of external/internal variables is just formal and  convenient for  easy calculations. 
Because of Poincar\'e  invariance and predictivity, $V$ can be only a function of 
the five independent scalars~\cite{annihp}
\beq   P^2 ,  {\ztil }^2 ,  {\ytil}^2 ,    \ztil \cdot \ytil ,    y \cdot  P 
  \label{five}  \eeq
But in view of    (\ref{yP2surP2})  it is more convenient for  calculations to re-arrange them as
\beq  P^2 ,  {\ztil }^2 ,  {\ytil}^2 ,    \ztil \cdot \ytil ,   
     {( y \cdot  P  )^2    \over  P^2  }    \label{5scalars}           \eeq
Unless otherwise specified  we consider $V$ as a function of these five arguments.  
Soon it was observed \cite{annihp}           that  
$$(X_1 - X_2 ) \ztil =  (X_1 - X_2 ) \ytil = 0$$  
implying this
\beprop 
 In the motion,   $\ztil$ and  $\ytil $   will depend only  on  
$\tau _1  +  \tau _2$.
\enprop
\noi In contrast, $z \cdot P$ and  $Q \cdot  P$  will depend on both evolution parameters.

\bigskip

For any unipotential model, the  {\sl individual energies }
$ \disp   \      {P \cdot p_1 \over  |P|} ,  \qquad    {P \cdot p_2 \over  |P|} \     $
are separately conserved.   Moreover the 
translation invariance of  $V$ implies conservation of $P^2$, hence this obvious first integral \beq N = H_1 + H_ 2  - { (H_1 - H_2)^2  \over P^2 } - {P^2 \over 4}  
                                                              \label{defN}     \eeq
 By elementary manipulations    we find  
\beq      N  =  \ytil ^2  +   2 V            \label{Nytil}                  \eeq
This important function defined on phase space   is    intimately related with the  properties of relative motion; this can be intuitively seen  as follows:  

\noi  fixing numerical values to   $H_1 , H_2 , P^\alp      $
 (in particular $P \cdot P = k \cdot k =  M^2$)  results in a numerical value for $N$,   let it be
$$ <N> =  - \Lam$$
The quantum mechanical analog of this quantity appeared, denoted as  $b^2$,  in the work of Todorov~\cite{todmass}.

\noi 
Employing  (for instance)  the reduced mass of Galilean mechanics, say   
$$\    \disp  m_0 = {m_1 m_2  \over  m_1 +  m_2      }         $$
we can start from  (\ref{Nytil}) and  check that  $\   \Lam / 2 m_0 \    $ is the leading term in the post-Galilean development of the quantity  $M - (m_1 + m_2 )$  which is usually considered as     
 {\sl binding energy} for the bound states of relativistic quantum mechanic (see Appendix I).

\noi   In addition we shall see later on that $\half N $ generates the evolution of the spatial relative canonical variables according to a one-parameter Hamiltonian scheme reminiscent of the nonrelativistic one-body mechanics, see equations (\ref{evolztil})  (\ref{evolytil}) below.

\bigskip

 From  (\ref{defN}) and  with the notation   (\ref{munumass})   we can write   
\beq   \Lam =    {M^2  \over 4} +
  {\nu ^2   \over M ^2 }   - \mu   
                                    \label{defLamc}       \eeq 
This important relation between $\Lam$  and  $M$  can be solved for  $M^2$. We first write
\beq  M^4 - 4 (\mu  +  \Lam  )  M^2  + 4 \nu ^2  = 0 
                                                     \label{bicar}     \eeq
\noi  From  (\ref{defLamc}) it is already clear that   $\Lam + \mu    >0$. 
     But $M ^2$ must be real and positive, so the possible values of $\Lam$ are  further 
restricted by the condition
\beq    | \mu  +   \Lam  |  >   |\nu |      \label{M2reel}         \eeq
which ensures that $M^2$ is real;
under this condition we can write
\beq M^2 = 2 ( \mu   + \Lam   )  \pm  2 \  
   \sqrt{ (\mu  +  \Lam  )^2  - \nu ^2 }      \label{M2}          \eeq
Moreover we must have at least one root  of  (\ref{bicar}) positive. Since their product is non-negative these roots  cannot have opposite signs, thus it will be sufficient to ensure that their sum is positive. Hence  the condition 
\beq      \mu  +  \Lam      > 0          \label{M2positif}        \eeq
 which ensures $M^2 > 0$.   We can encompass  both (\ref{M2reel})  and   (\ref{M2positif})  by  writting
\beq               
            \mu  +   \Lam    >   |\nu |       \label{reelposit}
\eeq

\medskip
\noi       Fortunately, the sign ambiguity in   (\ref{M2}) can be removed, with help of the individual     positive-energy condition (remind that $P \cdot p_a  / M$ are the {\em individual} energies) that we assume henceforth
$$  P \cdot p_1 >0,  \qquad       P \cdot p_2 >0             $$ 
Indeed, the numerical values of  $P \cdot p_1 $ and  $P \cdot p_2$ are given by
$ \disp  { M^2   \over    2 }    \pm  \nu  $.
 Requiring that both are {\em strictly  positive}  amounts to the condition
  \beq   M^2 >  2 |\nu |    \label{energposits} \eeq
  Let  ${M'}^2 ,  {M''}^2 $
 be the roots of equation   (\ref{bicar}). If both roots were to satisfy this 
{\em inequality}, it would contradict the {\em  equality} 
 $ {M'}^2   {M''}^2  =  4  \nu ^2 $ 
implied by the last term   in (\ref{bicar}). It follows that

\beprop  Under the condition of positive individual energies, only one
 root of    (\ref{bicar})   is admissible            \enprop 
 
\noi ( Remark $\   $For strictly equal masses, only the plus sign may be taken in (\ref{M2}).
Indeed this case means  $\nu = 0$, one root of equation (\ref{bicar}) is obviously zero which must be rejected, and the other root is   
$ M^2 =  4 ( \mu +  \Lam  ) $ obtained from   (\ref{M2})  by  choosing the  {\em plus}  sign).

We can  assume that  $m_1 \leq m_2$ without loss of generality. 
Thus  $\nu  \leq 0$, and  (\ref{reelposit})    becomes 
\beq            \mu + \nu + {\Lam }  > 0      \label{munuLam}     \eeq                         or equivalently
\beq    m_1 ^2 +  {\Lam  } >  0              \label{boundLam}  \eeq 
 We see that either  $\Lam >0$  or  it satisfies 
$ \disp  - m_1^2  \   <  \Lam \   < 0  \    $. In other words,

\beprop
 Under the assumption of positive individual energies, either  $\Lam  > 0$    or  $|\Lam | <  m_1 ^2$.
\enprop

\noi
If we impose the individual energy conditions we additionally get 
$ \disp  \half M^2  > \nu $ (trivial) and also     $ \disp  \half M^2  >   - \nu $, in other words
 \beq  M^2   >    m_2 ^2 - m_1 ^2         \label{boundM}           \eeq
Thus, as soon as  $m_1 \not=  m_2$,  the collective mass of the whole system (lenght of linear momentum) cannot be   arbitrarily small.

\bigskip
\noi   Let us check that taking the  {\em plus sign} in   (\ref{M2}) {\em always} yields an admissible root; eqn (\ref{munuLam}) implies that 
  $\mu + {\Lam }  >   - \nu    $, thus  looking at (\ref{M2})   we can  write
$$ {M^2 \over 2} >   - \nu   +     \sqrt{ (\mu  +  \Lam  )^2  - \nu ^2 }   
>   -  \nu                        $$
  implying the individual energy condition   (\ref{energposits}). []

  \medskip
\noi In view of the Proposition  2   
 above, taking the  minus sign in   (\ref{M2})is excluded  and we can write
\beq M^2 = 2 ( \mu + \Lam  )   + 
 2 \   \sqrt{ (\mu + \Lam  )^2  - \nu ^2 }      \label{M2+}        \eeq


\noi Note that $M$ reduces to $ m_1 + m_2 $ in the nonrelativistic limit, see Appendix~I.

%



\bigskip
We now investigate  the  evolution of the dynamical variables.
\noi Let us first analyze the evolution of  $\ztil$ and $\ytil$.
\beq  (X_1  +  X_2) \       \ztil ^\alp =  
     (X_1  + X_2)_{\rm free} \     \ztil ^\alp    +
  2  \{   \ztil ^\alp  ,  V  \}              \label{evolztil}    \eeq    
 
\beq  (X_1  +  X_2) \       \ytil ^\alp =  
     (X_1  + X_2)_{\rm free} \      \ytil ^\alp    +
  2  \{   \ytil ^\alp  ,  V  \}              \label{evolytil}    \eeq    
 where $V$ is a function of  the quantities listed in   (\ref{5scalars}).

\beprop        The  Poisson brackets
$ \{   \ztil ^\alp  ,  V  \} $ and $ \{   \ytil ^\alp  ,  V  \} $ are combinations of   
$\ztil ^\alp , \     \ytil  ^\alp$   with coefficients that are     functions of  
the five scalars listed in  (\ref{5scalars}).
\enprop

\noi Proof. 

\noi Consider first $\ztil ^\alp$. Obviously 
$ \disp  \{   \ztil ^\alp  ,  P^2   \} =    \{   \ztil ^\alp  ,   \ztil ^2   \}  =0    $,    and we  also have that  
    $ \disp   \{   \ztil ^\alp  ,  y  \cdot  P   \} =  0   $. 

\noi
Then we compute 
$$   \{   \ztil ^\alp  ,  \ytil ^2   \} =  2  \ytil ^\alp         $$
$$   \{   \ztil ^\alp  ,  \ztil  \cdot  \ytil    \} =    \ztil ^\alp         $$
   hence
\beq
  \{   \ztil ^\alp  ,  V  \}  =
 2   {\dron V  \over \dron  \ytil ^2 } \  \ytil ^\alp   +
     {\dron V  \over \dron  (\ztil  \cdot  \ytil ) }  \     \ztil  ^\alp
\label{ztilV}   \eeq
Since $V$  is a function of the scalars    (\ref{5scalars}) only,           
its partial derivatives involved in       (\ref{ztilV})     obviously share this property.

\noi   Then consider          $\ytil ^\alp$. Obviously  
$$  \{   \ytil ^\alp  ,  P^2   \} =    \{   \ytil ^\alp  ,   \ytil ^2   \} 
   =   \{ \ytil ^\alp  , \     {( y \cdot P )^2    \over   P^2 } \}  =  0  $$

\noi
Then we compute 
$$   \{   \ytil ^\alp  ,  \ztil ^2   \} = - 2  \ztil ^\alp         $$
$$   \{   \ytil ^\alp  ,  \ztil  \cdot  \ytil    \} =   - \ytil ^\alp         $$
  hence
\beq
  \{   \ytil ^\alp  ,  V  \}  =
  -2   {\dron V  \over \dron  \ztil ^2 } \  \ztil ^\alp   -
     {\dron V  \over \dron  (\ztil  \cdot  \ytil ) }  \     \ytil  ^\alp
\label{ytilV}   \eeq
The partial derivatives involved in   this formula  are functions of the scalars (\ref{5scalars}).~[]

\medskip   \noi 
In view of  Prop. 1  
 it is convenient  to  set
\beq    \lam = \tau_1 + \tau_2                            \eeq
the equations of motion for  $ \ztil ,  \ytil $  are
  \beq 
 {d \ztil  \over  d \lam  } =   \{ \ztil ,  \half    \ytil ^2   + V  \} 
                  \label{evolztil}         \eeq
 \beq 
 {d \ytil  \over  d \lam  } =   \{ \ytil ,  \half    \ytil ^2   + V  \} 
                    \label{evolytil}                 \eeq
where the brackets can be computed as  functions of   $ \ztil ^\mu , \ytil ^\nu ,  $   and  of  the first integrals  $P^2 ,   \     y \cdot P$.
 Once  $P^2$ and $y \cdot P $ have been fixed,
the evolution of the  spatial internal variables is given by a system of six 
first-order differential equations, to solve for six unknown functions; this problem has the structure of a {\em nonrelativistic}  problem  for one body in  three dimensions. 
The four-vectors    $\ztil$ and     $\ytil$ remain within the 2-plane  orthogonal to $k$ and to the (conserved) Pauli-Lubanski vector~\cite{annihp}.           
Some solution    
$$ \    \ztil =  \zeta (\lam, P^2 ,  y\cdot P ) , \qquad
        \ytil =  \eta (\lam , P^2 ,  y\cdot P)   \     $$
 of the system (\ref{evolztil})  (\ref{evolytil}) 
 defines  the evolution  of the spatial relative canonical variables $\ztil ,  \ytil $. Interpretation in terms of world lines,  relative positions and relative orbit will be given in the next section.

\noi The collective evolution parameter  $\lam$  plays the role of the Newtonian time  in the analogous one-body  system.    But in general $\lam$ {\em is not} the time of any inertial observer. Therefore, in order to evaluate the  {\em  schedule } of  the relative motion, we should express $\lam$ in function of     $T$  and insert 
the outcome into $\zeta $.


\noi A complete knowledge of the motion also requires that we  determine  the  evolution of 
 $z \cdot P$  and  $Q \cdot P$ in terms of  $ \tau_1 ,  \tau_2 $.

\noi
By sum and difference  
\beq    (X_1  +  X_2) \       z  \cdot  P  = 
        2y  \cdot  P  +    
        2    \{ z  \cdot  P ,    V   \}         \eeq    
\beq    (X_1  -  X_2) \       z  \cdot  P  =   P ^2                 \eeq  

We know that $V$ depends only on the five scalars   (\ref{5scalars}) and we observe that     $ \{  z\cdot P ,   \ytil _\alp \} = 0$, implying that 
 $  z\cdot P $ has a vanishing Poisson bracket with all scalars (\ref{5scalars})  except  $(y  \cdot  P )^2  /      P^2 )$. We find  
$\{ z\cdot P , \   y  \cdot  P   \}  =   P^2 $ hence finally

\noi   $  \{ z \cdot P , V  \}  $ {\sl only depends on the five scalars }(\ref{5scalars}).


After integrating the system 
(\ref{evolztil})(\ref{evolytil})   
let   us   set  
   \beq  \{ z  \cdot  P ,    V   \}   =    G ( \lam ,  M^2  ,  \nu  )  
\label{defG}      \eeq
   We obtain
$$( X_1 + X_2 )   z \cdot P =  2 \nu  +  2  G (\lam , M^2  , \nu )   $$
$$ ( X_1 - X_2 )   z \cdot P =    M^2                                 $$
Since  
$$  ( X_1 + X_2 ) =  {\dron \over  \dron \tau_1} + {\dron \over  \dron \tau_2} =
  2  {\dron \over  \dron  \lam }   $$
$$  ( X_1 - X_2 ) =  {\dron \over  \dron \tau_1} - {\dron \over  \dron \tau_2} =
  2  {\dron \over  \dron (\tau_1 - \tau _2 ) }   $$  
we  finally   have 

\beq  z \cdot P = \nu \lam   +   \int  G  d \lam  + 
 {M^2 \over 2}\    (\tau _1  - \tau _2 ) 
            +  {\rm const. }         \label{evolzP}  \eeq
Observing that
$$ {M^2 \over 2}  \pm  \nu  =  P \cdot p_1 \     ({\rm resp.}     P \cdot p_2 )    $$
we may write equivalently
 \beq  z \cdot P =   ( P \cdot p_1 )       \tau _1   
                     - (P \cdot p_2 )      \tau _2
 +   \int  G  d \lam      +  {\rm const. }
\eeq 
which reduces to eq. (3.6) of \cite{annihp}  when $G$ vanishes.

\noi Similarly   in view of  (\ref{yP2surP2})   and   (\ref{QPzytil})
we  can  simply write  (with $\dron$ according  to  (\ref{5scalars}) )
$$
\{ Q \cdot P , V  \} =   { \dron V  \over  \dron  P^2 }    \{ Q \cdot P ,  P^2  \}
$$
where  $\disp     \{ Q \cdot P ,  P^2  \}  =  2  P^2   $  therefore 
\beq   \{ Q \cdot P , V  \}  =   2  P^2 \     { \dron V  \over  \dron  P^2 } 
              \label{QPV}\eeq       
which  {\sl only depends on the five scalars }(\ref{5scalars}). 

\noi After integration of the system   (\ref{evolztil})(\ref{evolytil})           let us set   
\beq     \{ Q \cdot P , V  \} =  F ( \lam , M^2 , \nu )     
\label{defF}           \eeq
in other words we perform the substitution
\beq       F    =
 {\rm subs.} \   ( \ztil = \zeta  ,\    \ytil = \eta  , \   
         P^2 = M^2 , \    y \cdot P = \nu  |   \quad    \{ Q \cdot P , V  \} \    )                      \label{Subs}\eeq

\bigskip
\noi      Now  we can write 
\beq     (X_1 - X_2 )  Q \cdot P  = \{ Q \cdot P ,  y \cdot P \} =  y \cdot P 
                         \label{520}  \eeq
\beq     (X_1 + X_2 )  Q \cdot P  =
 (X_1 + X_2 ) _{\rm free }  +   2  \{ Q \cdot P , V  \}                      \eeq
\beq     (X_1 + X_2 )  Q \cdot P  =
                     \half P^2  +   2  \{ Q \cdot P , V  \}        
                         \label{522}          \eeq
Straightforward integration yields
\beq           Q \cdot P  =  {\nu \over 2} \    (\tau_1 - \tau_2 ) 
 + { M^2   \over 4 }  \lam 
         +      \int F  d\lam   + {\rm const.}         \label{evolQP}     \eeq    
We see that {\em modulo the solving of  (\ref{evolztil}) (\ref{evolytil})}                     the evolution of $Q \cdot P$ in terms of 
$\tau _1 , \tau_2$ will be given by a quadrature.
In the {\em very special case} where  $F \equiv 0$ (with our present  notation) the above formula 
  reduces to  eq (3.8) of  \cite{annihp}.
But most realistic potentials actually depend on $P^2$ which implies that $F$ differs from zero.

In contrast $G$ vanishes in several cases of interest, for instance

\noi {\sl  $G \equiv 0$ provided
  $V$ depends only on the dynamical variables  $\ztil ^2 ,  P^2 ,  L^2 $ }.  

\noi This statement stems from  (\ref{esptmp}).

\bigskip
To summarize:  After integrating  (\ref{evolztil})(\ref{evolytil}) we got $z \cdot P$ and $Q \cdot P$  as 
functions of $\tau _1 , \tau_2$.  Since we have the first integrals
$ P^\alp = k^\alp$  and  $y \cdot P  =  \nu$ the only remaining dynamical variables to be determined are $ {\widetilde Q }^\beta $. According to   (\ref{Qtil}) 
$${\widetilde {Q }} = {z \cdot P \over  P^2 }\   \ytil  -
                    {y \cdot P \over  P^2 }\   \ztil  -  
{  \widetilde{   P \cdot  M } \over P^2  }                        $$
where the last term is also a first integral, thus  everything in the  right-hand side is already detrmined.  This observation  achieves to determine the evolution of $q_1 ,q_2 , p_1  , p_2$, say
\beq  q_a =  \phi _a (\tau _1 , \tau _2 ) ,   \qquad             
      p_b =  \psi _b (\tau _1 , \tau _2 ) ,   \qquad           \label{evolqp}  \eeq
 These functions define   
 the  two-dimensional "orbits" \footnote{To avoid confusion with trajectories in space, we put the word {\em orbit} between quotation marks when it is  meant in the group-theoretical sense.} 
 of the evolution group  in phase space.

\section{World lines in Unipotential Models}	 
The solutions of system (\ref{hamil})  can  be interpreted in terms of world lines provided we ultimately  introduce the physical coordonates $x_1 ^\alp, x_2 ^\beta$ as functions of the canonical coordonates.
Our Cauchy surface for solving the position equations (\ref{positeq}) is $(\Sig )$ defined  by   $\   P \cdot  z  =0 \   $, and our initial condition 
\beq     x_a ^\alp  -  q _a ^\alp  = 0  \qquad  \quad   {\rm on \        the \   
  surface}  \qquad      ( \Sig )           \label{prescrip}          \eeq  
can be formulated also as
\beq   x_a ^\alp =  q_a ^\alp  +  O (P \cdot  z )                  \eeq
where  $O (P \cdot  z ) $ symbolically represents any expression which vanishes with 
$P \cdot  z$.  
In principle   formula (\ref{evolqp})  must be inserted into the solutions of  (\ref{positeq}) and  yields the worldlines in terms of the individual evolution parameters, say
$\   x_1 (\tau_1 ),  x_2 (\tau_2 )\   $.

\noi The above  prescription 
 offers  several advantages. First of all, setting 
 $$r^\alp  = x_1 ^\alp -  x_2 ^\alp  $$
 condition   (\ref{prescrip}) implies that also 
 $P \cdot r$ vanishes on  $(\Sig )$. In the rest frame we can write 
$\   x_1 ^0 =  x_2 ^0 = T,\   $
 thus  finally the manifold $(\Sig ) $ can be called the {\sl Equal-Time Surface }.

\noi {\em  At equal times} the radius-vector  ${\widetilde r} ^\alp $  moves on a curve that we may call the {\sl relative orbit}.
 As observed long time ago~\cite{annihp},  this curve lies on the 2-plane mentioned in the previous section (orbital plane).

  This version of  the Hamiltonian formalism 
clarifies the formal definition (\ref{vardefcdm}) written  for the center of mass.
  
\noi Indeed  (\ref{vardefcdm})  is equivalent to (\ref{varvarcdm}), say  

$$ \Xi =   {{ (P \cdot p_1  )  \   q_1  +   ( P \cdot p_2  ) \    q_2  } \over  P^2  } 
  + O ( P \cdot  z ) 
$$
In terms of the {\em individual energies } $M_a = ( P \cdot p_a ) /  |P| $, we have
\beq \Xi =  { M_1  q_1 + M_2 q_2  \over  M_1 + M_2 }  + O ( P \cdot  z )
\label{sympacdm}   \eeq
Now {\em at equal times } $P \cdot  z$ vanishes; fixing $P^\alp = k^\alp$  
we can replace $q_a$ by $x_a$ and  $P^2 $ by $M^2$, so we are left with
\beq \Xi |_\Sig =  { M_1  x_1 + M_2 x_2  \over  M_1 + M_2 }      \label{physcdm} 
 \eeq


Notice that $M_1 + M_2 = M$ and remember  
 that  the individual energies
reduce to the masses in the nonrelativistic limit (see Appendix II).
In view of these remarks, formula (\ref{physcdm}) is more intuitive and significant  when  
$P \cdot p_1 $  and      $ P \cdot p_2$ are both positive:
the analogy with the Newtonian  definition of center of mass becomes obvious,  which legitimates the positive-energy condition. 
At this stage it is clear that definition (\ref{vardefcdm}) agrees with  the one  proposed  by  Pryce~\cite{pryce}; similarly,  formula (\ref{physcdm})  agrees with the notion of {\sl center of energy} according to    Fischbach {\em et al} \cite{fisch}.

\subsection{Rest-Frame description}

In practice, instead of trying to describe the motion in terms of the independent parameters  $\tau _1 , \tau _1$ we have better to fix  the  linear momentum  $k^\alp$, so  defining a slicing of spacetime by the  three-planes orthogonal to $k^\alp$.
These three-planes intersect both world-lines, which provides the rest-frame description of dynamics, as follows:  

\noi  among all possible couples  $\   x _1 , x _1 \   $  
 the slicing  selects the equal-time configurations, characterized by    $k \cdot r = 0$.
 We just have to determine the sequence of these configurations, that is a {\em one-parameter set}. Picking up the equal-time configurations  obviously induces a relation between    
$\tau _1 $ and $ \tau _2$,  by cancellation of   $\   P \cdot  z   \   $ in formula
(\ref{evolzP}). One is left with a (possibly nonlinear) expression of 
$\tau _1  - \tau _2$ as a function of $\lam$.

 \bigskip


\noi After imposing  (\ref{prescrip}),  formula  (\ref{evolzP}) permits us to express  everything    in terms of $\lam$ only. 
Formula  (\ref{vardefcdm}) permits to write
\beq  q_1 = \Xi  -  ( {\nu \over M^2} -  \half ) \      z  +  O (P \cdot z )   
\label{q1sig} \eeq 

\beq  q_2 = \Xi  -  ( {\nu \over M^2} +  \half )  \      z   +  O (P \cdot z )   
\label{q2sig} \eeq 
But  $ z = \ztil   +  O (P \cdot z ) $.    
Taking  (\ref{prescrip}) into account 
(which implies      $\ztil =  \rtil  +  O (P \cdot z )    $),  we put  $P \cdot z$ equal to zero in (\ref{q1sig}) (\ref{q2sig})   and   obtain the {\sl equal-time description of the motion}
\beq  x_1 = \Xi  -  ( {\nu \over M^2} -  \half )  \zeta (\lam , M^2 , \nu )
  \label{x1sig}  \eeq 
\beq  x_2 = \Xi  -  ( {\nu \over M^2} +  \half )    \zeta (\lam , M^2 , \nu )          
  \label{x2sig}  \eeq 
These formulas yield a representation of both  world lines in terms of the same parameter $\lam$.
But for the sake of a better understanding of the motion
 it  is interesting  to express $\lam$  as a function of the center-of-mass proper time  $T$.
 
\noi Cancelling  $P \cdot z$    in    (\ref{evolzP}),  our  equal-time prescription implies
$$ \half M^2 (\tau _1  - \tau_2 )  = - \nu \lam  -  \int G d \lam + {\rm const.} $$
Inserting into   (\ref{evolQP}) yields the common value of  $\Xi ^0$ and $Q^0$ in the center-of-mass frame, say  

\beq   T  =    \lam ( {M \over 4}  -  {\nu ^2  \over M^3 } )   
-  {\nu \over  M^3}  \    \int G d\lam
        + {1 \over M}     \int F d\lam     + {\rm const.}                              \label{tempcdm} \eeq
In principle this equation  must be solved for  $\lam$.  Let us stress that
{\em only in the very special case 
 where $G$ and $F$ are constant},  the time of the center of mass  is {\em for all orbits } a linear function of the parameter    $\  \lam \  $.   

\noi For instance,  this situation is realized when     both 
  $\dron V  / \dron P^2$    and
 $\dron V / \dron  (y \cdot P ) $ identically vanish, implying  $G =F =0$.
This situation will be referred to as the    {\sl  Academic Case}. 

\noi Otherwise, in the most general case $\lam$ and $T$ are related in a  nonlinear  way,  but  for exceptional orbits.

\medskip
Note that for a  physically admissible solution to  (\ref{evolztil}) (\ref{evolytil}),   $T$ should   monotonously  increase as a function of $\lam$.
In the academic case this is automatically ensured by the condition 
(\ref{energposits}) of positive individual energies.
 When the interaction potential $V$ is more complicated we must demand  
 $d T / d \lam > 0$, whereas we can write
\beq  {d T \over d \lam}  =
{M \over 4}  -  {\nu ^2  \over M^3 } -  {\nu G \over M^3} +  {F \over M}  
\eeq 
For example assume for a moment that $G \equiv 0$, we are left with a simpler condition. If $F$ is positive no problem; otherwise the positive-energy condition must be replaced by a more restrictive and model-dependent condition 
({e.g.} see the toy model of Section 6 ).
 
\noi More  generally, the discussion remains easy when $F$ and $G$ are bounded; as we shall see in the following section, this circumstance arises in  case of circular motion.

\subsection{Circular Motion} 
Circular motion is characterized by the constancy of $\ztil ^2$, which implies that {\em at equal times}  ${\widetilde r} ^2$  also is constant.
We can check that

\beprop On any circular orbit the functions $G, F$ and the five dynamical variables 
(\ref{5scalars}) are constant.  
\enprop

\noi Proof.

\noi The interaction potential is a function of the five dynamical variables
(\ref{five}) 
We have four remarkable constants of the motion
$ P^2 , N , y \cdot P $  and the square of angular momentum
\beq  L^2  =  \ztil ^2   \ytil ^2    -  (\ztil   \cdot    \ytil ) ^2     
                                  \label{Luban}   \eeq
when fixed, they respectively take on the following numerical values
$$   M^2 , - \Lam ,  \nu   ,  l^2       $$
In the set (\ref{five}) we can replace     
$  \ztil ^2  ,  \ytil ^2 ,  \ztil \cdot \ytil  $   by  the equivalent  set of scalars
$  \ztil ^2  ,  \ytil ^2 ,   L^2    $. So let
$$ V  =  f (P^2 ,  \ztil ^2  ,  \ytil ^2 ,   L^2    ,  y \cdot P   )     $$
Since  $N = \ytil ^2  +  2V$   we can write
$$ N - \ytil ^2 =  2   f (P^2 ,  \ztil ^2  ,  \ytil ^2 ,   L^2    ,  y \cdot P   )    $$
This equation implicitly defines  $\ytil ^2  $ as a function of 
$P^2 , \ztil ^2 , N , L^2 ,    y \cdot P  $, if we leave apart a very exceptional case where $V$ linearly depends on $\ytil ^2$ in a special manner (such case is not realistic anyway).

\noi  Therefore it is sufficient that  $\ztil ^2 = \const \   $ for having  also
   $\   \ytil ^2 = \const \   $,  which in turn,  according to  (\ref{Luban}), implies   $\ztil \cdot \ytil  = \const$.

\noi Finally the five dynamical variables    (\ref{five}), or equivalently  (\ref{5scalars}),     remain constant on the circular orbit.  []   

It follows that $\lam$ is a linear function of $T$ on circular orbits.

\betheo
If the interaction  is such that  $ \{ \ztil , V \} = 0$   
 and  $ \disp   {\dron  V \over \dron (\ztil \cdot \ytil ) }   = 0$
     there  exist  circular orbits; on these orbits  the relative motion is periodic in terms of the center-of-mass time.
\entheo

\noi Proof. $\   $   According to (\ref{evolztil}) we have that   
$ \disp  {d \ztil\over  d  \lam} =  \ytil   $  and acording to    (\ref{evolytil})
we get
 $ \disp   {d \ytil \over  d  \lam} =    - 2  {\dron V \over  \dron  \ztil ^2 } \    \ztil$.
Equations (\ref{evolztil})(\ref{evolytil}) are similar to that of a three-dimensional two-body problem. The case when  $ \{\ztil ^\alp    ,\   V     \}$  is zero corresponds to the classical problem of motion under a central force, where circular orbits are known to exist.

\noi  As seen above,    $\ytil ^2 $ is  constant.  In the analogy between  our system and that of Galilean  mechanics, $\lam$ plays the role of time and $ - \ytil ^2 $ represents the squared velocity.  As well as a circular motion with a velocity of constant lenght is necessarily periodic in time, 
here  we have that   $\ztil $  and    $\ytil $   are  periodic functions of  $\lam$.
Since we consider circular motion, $T$ is a linear function of  $\lam$ and 
{\em vice versa},  so periodicity in $\lam$  implies periodicity in $T$.  []


\noi Example:  any  $V(P^2 , \ztil ^2 )$ admits   circular orbits.


\section{Extreme mass ratio, one-body limit }

\noi We keep considering unipotential models.
The case where   one  mass can be neglected in front of the other one is of practical interest when one tries to  justify  a resonable  expression  of  $V$. Indeed it is naturally  expected   that in the  limit of an extreme  mass ratio  we  recover a system made of particle $1$ moving in the external  field created by particle $2$,  the latter undergoing rectilinear uniform motion.

Without loss of generality we assume $m_1   \ll  m_2$. Note that we cannot  just  put  $m_2$ to infinity.
This can be seen already  in the framework of Newtonian mechanics, because  the gravitational potential  created   around  particle $2$  could not  remain  finite  when  $m_2 $ tends to infinity.
Therefore we shall rather put
$$  \    m_1   =  \gam \      m_2 \    $$  
         and study  the limit  for     $  \gam  \rightarrow   0$.

\noi   In order to alleviate calculations, let us  set   $\    \vareps = \gam ^2 \   $.
 We have 
\beq   \mu =  \half m_2^2 (\vareps + 1 )    \label{devlopmu}              \eeq
\beq   \nu =  \half  m_2 ^2 (\vareps - 1 )  \label{devlopnu}              \eeq            
whence we derive
$$   \mu ^2  -  \nu  ^2  =    {1 \over 4} \     m_2 ^4  \
    [ (\vareps + 1 )^2  - (\vareps - 1 ) ^2 ]    $$
 \beq
 \mu ^2  -  \nu  ^2  =  \vareps  \      m_2 ^4      \label{mu2nu2}  \eeq

\bigskip

\noi We whish  to investigate whether,  looking at things from the rest frame, the center of mass $\Xi$ and the spacetime position   $x_2$ of the most heavy body  actually   coincide  in the  limit   $\gam \rightarrow 0$.
By formula  (\ref{vardefcdm})  we may write 
\beq   \Xi =  Q  +  ( { y \cdot P  \over P^2 }  )  \    z  
  -  ( { P \cdot z   \over  P^2 } ) \      y
 \eeq
where $Q = q_2  +   \half z   =  q_1  -  \half  z  $.  
On the mass shell we have that $P^2 = M^2$ so
\beq
 \Xi = q_2  +  ( \half   +   { \nu  \over  M^2 } ) z       
-  {P \cdot z \over M^2 } y 
                  \label{eobcdm}             \eeq


\noi  We can write   
     $\   m_1 ^2 =   \vareps \      m_2 ^2 \   $,  where, of course  
$\quad      \vareps =   \gam ^2$.

\noi We are interested in  what happens  when    $\gam \rightarrow 0$.


\noi In order to  consider the most general case,   let us set   
\beq   \Lam  =   \alp \    m_2 ^2  \label{alfaeob}    \eeq
 without assuming  for the moment any restriction about the 
magnitude~\footnote{Here we change the notation of \cite{cras} by suppression of a factor $\half$.}
 of  $\alp$.

In view of     (\ref{devlopmu})  and  (\ref{mu2nu2}) we get
$$ ( \mu + \Lam )^2  -  \nu ^2   =   m_2 ^2   \alp  \      (\Lam + 2 \mu )
                                   +  \vareps   m_2 ^4       $$
$$   (\mu + \Lam )^2  -  \nu ^2   =
      m_2 ^2  \     \alp   \     
  [  m_2 ^2   \alp  +   m_2 ^2    (\vareps + 1 )  ]      +  \vareps   m_2 ^4  
$$ 
\beq                  (\mu + \Lam )^2  -  \nu ^2   =
       m_2  ^4  \     ( { \alp  } + 1 )  (  { \alp  } + \vareps )
\eeq
Notice  that  $2 ( \mu + \Lam ) = m_2 ^2 (1 + \vareps + 2 \alp )$. 
Inserting into             (\ref{M2+})        yields     the rigorous formula
                \beq
M^2 =  m_2 ^2 [ 1 + 2 \alp  +  \vareps     +
   2  \sqrt{ (1  +  \alp ) ( \alp +  \vareps ) } ]
         \label{Malfaeps}                \eeq
 valid irrespective of the order of magnitude of $\alp$ and  $\vareps$.

\noi  Now we are in a position to make the following statement

 \betheo 
Provided  we can neglect        
$\   \sqrt{|\Lam |}  \   $    in  front of     $\   m_2  \    $, 
 we  have  that   $M^2   \rightarrow  -  2 \nu$, which entails that, 
{\em at equal times},
 $\Xi$ and $x_2$ coincide in the limit $\gam \rightarrow 0$.
\entheo
Indeed neglecting   $ \disp  {\sqrt{|\Lam |}  \over   m_2 } \    $ amounts to cancel  $\alp$ in    (\ref{Malfaeps}),   that yields 
 $ - \half$ as the limit of the ratio  $\disp \nu  / M^2$, making the second term in  the right-hand side of (\ref{eobcdm})      to vanish; remember that the third term vanishes at equal times.   []

\bigskip
\noi   Owing to Propo. 3  the condition for this result is always satisfied  for  negative $\Lam$.
 
\noi In contradistinction     large positive values of $\Lam $ forbid $\Xi$ to coincide with  the heavy body in the limit of an extreme mass ratio.
  In this case $\alp > 0 $  and   
  formula  (\ref{Malfaeps}) can be written
$$  M^2 =  m_2 ^2 \    [  1 +  2 \alp  + \vareps   +  2  \sqrt{(1 + \alp )  \alp } \quad       \sqrt { 1+   \vareps / \alp } ]    $$ 
Define       
\beq     \beta = 2 \alp   + 2 \sqrt{\alp^2 +   \alp }  \     \label{defbeta}    \eeq
 Since  $\alp > 0$  it is clear that $\beta > 2 \alp$. We get

\beq  M^2 =  m_2 ^2  (  1 + \beta  ) +   O ( \vareps )    \label{M2beta}           \eeq
now using     (\ref{devlopnu})                              yields
 \beq     \half +  {\nu \over M^2 }  =   {\beta \over  2 (1 + \beta ) }       +         O ( \vareps  )          \eeq  
which in general {\em cannot  vanish}  when $\vareps  \rightarrow 0$. []

\medskip
\noi This situation can be physically interpreted as follows: Considered at equal times, our covariant definition of the center of mass $\Xi$ reduces to  that of    Fokker and Pryce \cite{pryce}. See also Moeller \cite{moell}.  Accordingly we  notice that
   $\   \Xi \   $ is in fact  a   {\em center of energy}; therefore 
 {\em not only the masses  but also the energies} must be taken into account.
 Even if  $m_1$ is very small, it must be understood that we cannot consider particle  $1$ as a test particle when its  motion involves a too large  amount of energy.

\section{Toy Model}.
Consider a harmonic potential 
\beq V = \chi   \   \sqrt{ P^2}  \ztil ^2       \label{oscar}      \eeq
with $\chi$ a positive string  constant (this potential differs from the one 
considered in \cite{annihp} by because it is $P^2$-dependent, which alows for the correct dimension of the coupling constant).

\noi The structure of the calculations derived from (\ref{oscar}) is that of a nonrelativistic problem. Our notation is such that the relativistic potential and its nonrelativistic conterpart have  opposite signs, so  we have a positive $\Lam$.

\noi   We must compute $F$, as defined in (\ref{defF}), according to  (\ref{QPV}) and after solving the reduced equations of motion (\ref{evolztil})(\ref{evolytil}).
The solution to this system is
\beq
\ztil =  A  \   \sin (\Ome \lam + C) +  B  \   \cos (\Ome \lam + C)
\label{ztiloscar}   \eeq
\beq
\ytil = A  \Ome  \    \cos (\Ome \lam + C)  - B  \Ome \    \sin (\Ome \lam + C)
\label{ytiloscar}
\eeq                                             
where  $A,    B$ are  mutually orthogonal spacelike constant vectors (they span the orbital plane, their lenghts are the half-axes of an ellipse) and    $C$ is a scalar constant; moreover we have
$$    \Ome = \sqrt{2 \chi |P| }             $$
  Note that
  $$     \{ Q  \cdot  P  ,   V \} =  2 {\dron V  \over \dron P^2 } P^2 =
  \chi   \   \sqrt{ P^2}   \ztil ^2 
   = V$$
which is always negative.
It is clear that  $F$ will depend on $\lam$  only through $\ztil ^2$.    
Taking   (\ref{ztiloscar}) into account and
fixing  $P^\alp  =  k^\alp$ we are left with   
$$    \Ome = \sqrt{2 \chi M }             $$  
whence we derive
$$ <N>  = 2 \chi M (A^2 + B^2 ) = - 2 \chi M (\joka  + \jokb  )         $$
$$  \joka  + \jokb  =     { \Lam    \over  2  \chi  M}                  $$
setting  $\    A^2 = - \joka, \     B^2 = -  \jokb      \    $. We obtain
\beq  F = - \chi M [ \joka  \sin ^2 (\Ome \lam + C) +
                     \jokb  \cos ^2 (\Ome \lam + C) ]
\label{Foscar}    \eeq  
In order to calculate   $\int F d\lam$ we notice the primitive  

$$    \int ^\lam   (\joka  \sin ^2 (\Ome \lam + C)
               +   ( \jokb  \cos ^2 (\Ome \lam + C) )\   d\lam   =    $$
$$  
 {\joka + \jokb \over  2 }  \     \lam
+ {\jokb - \joka \over  2 \Ome}    \sin (\Ome \lam + C)  \cos (\Ome \lam + C)
+ {\rm const.} $$   

Finally we find
\beq
\int ^\lam  F d \lam = 
- \chi M \    [   {\joka + \jokb \over  2 }  \     \lam
 +    {\jokb - \joka \over  4  \Ome }   \sin (2\Ome \lam + 2C) ]   + {\rm const.}
\label{intF}     \eeq     
so there is a secular (linear) term plus a periodic correction; 
  in this particular example $\widetilde r$ is  a periodic function  of $\lam $. 

\noi Since  $z \cdot P$ has a vanishing bracket with $P^2$ and $\ztil ^2$ it is obvious that    $\{ z \cdot P , V \}$  hence $G$, vanishes.

Owing to     (\ref{energposits})       we are sure that 
$ \disp \    {M \over 4} -  {\nu ^2  \over M^3 } > 0 $. But the question is about   
$ \disp  \    { d T  \over  d \lam  }   \     $. 
It can be directly read off     (\ref{Foscar}) that   
$$   |F| \leq  \chi M      (\joka + \jokb )$$
so the condition for         $\disp   { d T  \over  d \lam  }   > 0$ is
$$ {M^2  \over 4} -{\nu ^2 \over M ^2}   >    |F|              $$
But   
$$      |F|  \leq    \chi M      (\joka + \jokb )$$
hence a { \em   sufficient   condition}
$$ {M^2  \over 4} -{\nu ^2 \over M ^2}   >   { \Lam \over 2}   $$

\section{Conclusion}
Center of mass and relative motion are well understood concepts for isolated two-body systems, provided interaction is of the unipotential type and physical positions are fixed by the equal-time prescription. A large part of our picture  can be made abstractly but most physical features show up in the light of the equal-time description.
 The main scheme was put forward many years ago \cite{repor}, but several important consequences are considered here for the first time. 

\noi In the present work we regard seriously the fact that the collective evolution parameter which arises in the reduced equations of motion (\ref{evolztil}) (\ref{evolytil})  is generally {\em not} a linear function of the center-of-mass time. This peculiarity (which doesnot concern  the geometry of the orbit) automatically affects the {\em schedule} of relative motion.

\noi This point led us to consider an important exception, the   case of circular orbits, where the relative motion is   periodic in $T$.

 Although the individual evolution parameters $\tau_1 , \tau _2$ are generally not affine on the world lines, our equal-time treatment offers the possibility to end up with a description in terms of the center-of-mass time.

On the other hand, examinating the case of an extreme mass ratio provides an illustration of the nature of center of mass in relativity. Indeed the result expressed in  Theorem 2 forces one  to interprete $\Xi$ as a center of {\em energy} rather than of mass; this is in agreement with an ancient  literature and with the spirit of relativity.

In this paper we considered relative motion essentially by analogy with a nonrelativistic one-body problem, naturally suggested by  (\ref{evolztil}) (\ref{evolytil}).
But of course the question as to know whether (and how)  a  ficticious
 {\em relativistic} one-body system can be invoked, is relevant and deserves a separate publication.


%
%
%
%
%

%

$$                                    $$

{APPENDIX}

\noi I. $\    $Reverting from $\mu , \nu$ to $m_1 , m_2$,   formula      (\ref{M2+})  implies this useful approximation
\beq  M = m_1 + m_2  +  {\Lam \over 2 m_0  c^2}   +  O (1/c^4 )
                 \label{approxM}           \eeq
At first sight the appearence of  $m_0$ in this formula seems to indicate that the non-relativistic expression for reduced mas goes over to the relativistic realm without  modification; 
 notice however that we could replace  $m_0$ by  any positive $m$    in this formula, provided that 
$  m = m_0 +  O (1/c^2 )  \   $. This remarks leaves open the possibility that the "good"  relativistic generalization of the  reduced mass may  coincide with  $m_0$ only  in the nonrelativistic limit, as happens for instance with  Todorov's~\cite{todmass}        reduced mass  $\disp   m_T = {m_1 m_2  \over  M}$. 
 
\bigskip

\noi II. $ \    $ Defining
$ \    M_a  =  P \cdot p_a  /  M  c^2  \     $ we have 
$$ M_1 +  M_2  =  M ,  \qquad  \quad  
   M_1 -  M_2  =  {2 y \cdot P  \over  M c^2}   =   {2 \nu  \over  M }       $$  
But  $2 \nu =  (m_1 ^2  -  m_2 ^2 )  c^2$  thus 
 $\disp    M_1 -  M_2  =  { m_1 ^2  -  m_2 ^2  \over  M}   $.

\noi According to  (\ref{approxM}) we have  
$ \disp   {1 \over  M}  =    {1 \over  m_1  +  m_2  }  +  O (1/c^2 )   $,  so  finally
$$   M_1 -  M_2  =  { m_1 ^2  -  m_2 ^2  \over m_1  +  m_2 }          +  O (1/c^2 )  = m_1  -  m_2         +  O (1/c^2 )       $$                             
Finally  $   M_a = m_a  +  O (1/c^2 )       $.

%

\end{document}